\documentclass{iaus}
\usepackage{graphicx}
\usepackage{natbib}
\usepackage{aas_macros}
\title[Massive SN Progenitors] 
{Around the Pair Instability Valley - Massive SN Progenitors}

\author[R. Waldman]   
{Roni Waldman}

\affiliation{Racah Institute of Physics, Hebrew University,
Jerusalem 91904, Israel \break email: waldman@cc.huji.ac.il\\[\affilskip]
}

\pubyear{2008}
\volume{252}  
\date{?? and in revised form ??}
 \jname{Proceedings Title IAU Symposium}
\editors{Licai Deng, K.L. Chan \& C. Chiosi, eds.}
\begin{document}

\maketitle

\begin{abstract}
The discovery of the extremely luminous supernova SN~2006gy,
possibly interpreted as a pair instability supernova, renewed the
interest in very massive stars. We explore the evolution of these
objects, which end their life as pair instability supernovae or as
core collapse supernovae with relatively massive iron cores, up to
about $3\,M_\odot$.

\keywords{stars: evolution, (stars:) supernovae: general}
\end{abstract}

\firstsection 
\section{Introduction}
The interest in the evolution of very massive stars (VMS), with
masses $\gtrsim 100\,M_\odot$, has recently been revived by the
discovery of SN~2006gy - the most luminous supernova ever recorded
\citep{Smith2007ApJ...666.1116S}. This object, having a luminosity
of $\sim 10$ times that of a typical core-collapse SN (CCSN), is
probably the first evidence of a pair instability SN (PISN)
\cite{Woosley2007Natur.450..390W}. PISN are massive stellar objects,
whose evolutionary path brings their center into a region in
thermodynamical phase space $(\rho \lesssim 10^6, T \gtrsim 10^9)$,
where thermal energy is converted into the production of
electron-positron pairs, thus resulting in loss of pressure and
hydrodynamic instability. This type of supernova was first suggested
40 years ago by \cite{Barkat1967PhysRevLett.18.379}, and since then
several works were carried out
\cite[e.g.][]{Fraley1968Ap&SS...2...96F, Ober1983A&A...119...61O,
Ober1983A&A...119...54E, Bond1984ApJ...280..825B,
Heger2002ApJ...567..532H, Hirschi2004A&A...425..649H,
Eldridge&Tout2004MNRAS.353...87E, Nomoto2005ASPC..332..374N},
however the overall interest in this topic has been relatively
small, mainly due to lack of observational data.

It was originally believed that stars massive enough to produce PISN
could only be found among population III stars with close to zero
metallicity ($Z \lesssim 10^{-4}$), and hence only at very high
redshift ($z \gtrsim 15$). More recently
\cite{Scannapieco2005ApJ...633.1031S} discussed the detectability of
PISN at redshift of $z \leq 6$, arguing that metal enrichment is a
local process, therefore metal-free star-forming pockets may be
found at such low redshifts. \cite{Langer2007A&A...475L..19L}
introduced the effect of rotation into studying this question
concluding that PISN could be produced by slow rotators of
metallicity $Z \lesssim Z_\odot/3$ at a rate of one in every 1000 SN
in the local universe. Furthermore, \cite{Smith2007ApJ...666.1116S}
point out, that mass loss rates in the local universe might be much
lower than previously thought, so that massive stars might be left
with enough mass to become PISN. This conclusion is also supported
by \cite{Yungelson2008A&A...477..223Y} who extensively discuss the
mass loss rates and fates of VMS. It is interesting to note, that
SN~2006gy took place in the nearby Universe. Following the discovery
of SN~2006gy, \cite{Umeda&Nomoto2008ApJ...673.1014U} addressed the
question of how much $^{56}Ni$ can be produced in massive CCSN,
while \cite{Heger&Woosley2008arXiv0803.3161H} computed the detailed
nucleosynthesis in these SNe.

The interest in VMS is further motivated by the discovery of
Ultraluminous X-ray Sources (ULX), which can be interpreted as
mass-accreting intermediate mass black holes (IMBH) with mass $\sim
(10^2 - 10^5)\,M_\odot$. One of the possible scenarios for IMBH
formation is by VMS formed by stellar mergers in compact globular
clusters \cite[see e.g.][and references
therein]{Yungelson2008A&A...477..223Y}. In this context,
\cite{Nakazato2006ApJ...645..519N, Nakazato2007ApJ...666.1140N}
studied the collapse of massive iron cores with $M \gtrsim
3\,M_\odot$. In their first paper they treat the fate of stars of
mass $\geq 300\,M_\odot$ which reach the photodisintegration
temperature ($\approx 6 \times 10^9 K$) after undergoing pair
instability. The entropy per baryon of these models at
photodisintegration is $s>16k_B$ compared with the classical
core-collapse SN with $s \sim 1k_B$. In the second paper they aim to
bridge this entropy gap, corresponding to core masses of $(3 -
30)\,M_\odot$ but claim that there is a lack of systematic
progenitor models for this range, hence they use synthetic initial
models for their calculations.

In this work we focus mostly on the mass range $M \lesssim
80\,M_\odot$ (He core mass $M_{He} \lesssim 36\,M_\odot$)
immediately below the range which enters the pair instability
region, and present a systematic picture of the resulting CCSN
progenitors.

\section{Method}

Since the mass loss rates of stars in this range are highly
uncertain, \cite[see e.g. discussion
by][]{Yungelson2008A&A...477..223Y}, we avoid dealing with this
question by following the example of
\cite{Heger2002ApJ...567..532H}, and modeling the evolution of
helium cores. Our helium core initial models are homogeneous
polytropes composed entirely of helium and metals, with metallicity
$Z \approx 0.015$, in the mass range $(8-160)\,M_\odot$ . The models
were then evolved to the helium zero age main sequence. In the
following we will refer to these models as ``He\emph{N}'' where
\emph{N} is the mass of the model. For comparison we evolved also a
few models of regular hydrogen stars, beginning from the zero-age
main sequence (ZAMS). We will refer to these models as ``M\emph{N}''
where \emph{N} is the mass of the model. All our models have no
mass-loss. We argue that as long as the mass loss rate is not so
high that it will cut into the He-core, the evolution after the
main-sequence phase will be virtually independent of the fate of the
hydrogen-rich envelope. We followed the evolution of each model
until the star is either completely disrupted (for the PISN case) or
Fe begins to photo-disintegrate (for the CCSN case).

We followed the evolution using the Lagrangian one dimensional Tycho
evolutionary code version 6.92 (with some modifications), publicly
available on the web \cite[the code is described
in][]{Young_Arnett2005ApJ}. Convection is treated using the well
known mixing length theory (MLT) with the Ledoux criterion. In the
MLT formulation of Tycho, the value of the mixing length parameter
fit to the Sun is $\alpha_{MLT} \approx 2.1$
\citep{Young_Arnett2005ApJ}, so we used a value of
$\alpha_{MLT} = 2$ in our calculations. The nuclear reaction rates
used by TYCHO are taken from the NON-SMOKER database as described in
\cite{Rauscher_Thielemann2000ADNDT..75....1R}.

\section{Results and Discussion}
\begin{figure}
\begin{minipage}[t]{0.5\linewidth}
\centering
\includegraphics[width=1.\linewidth]{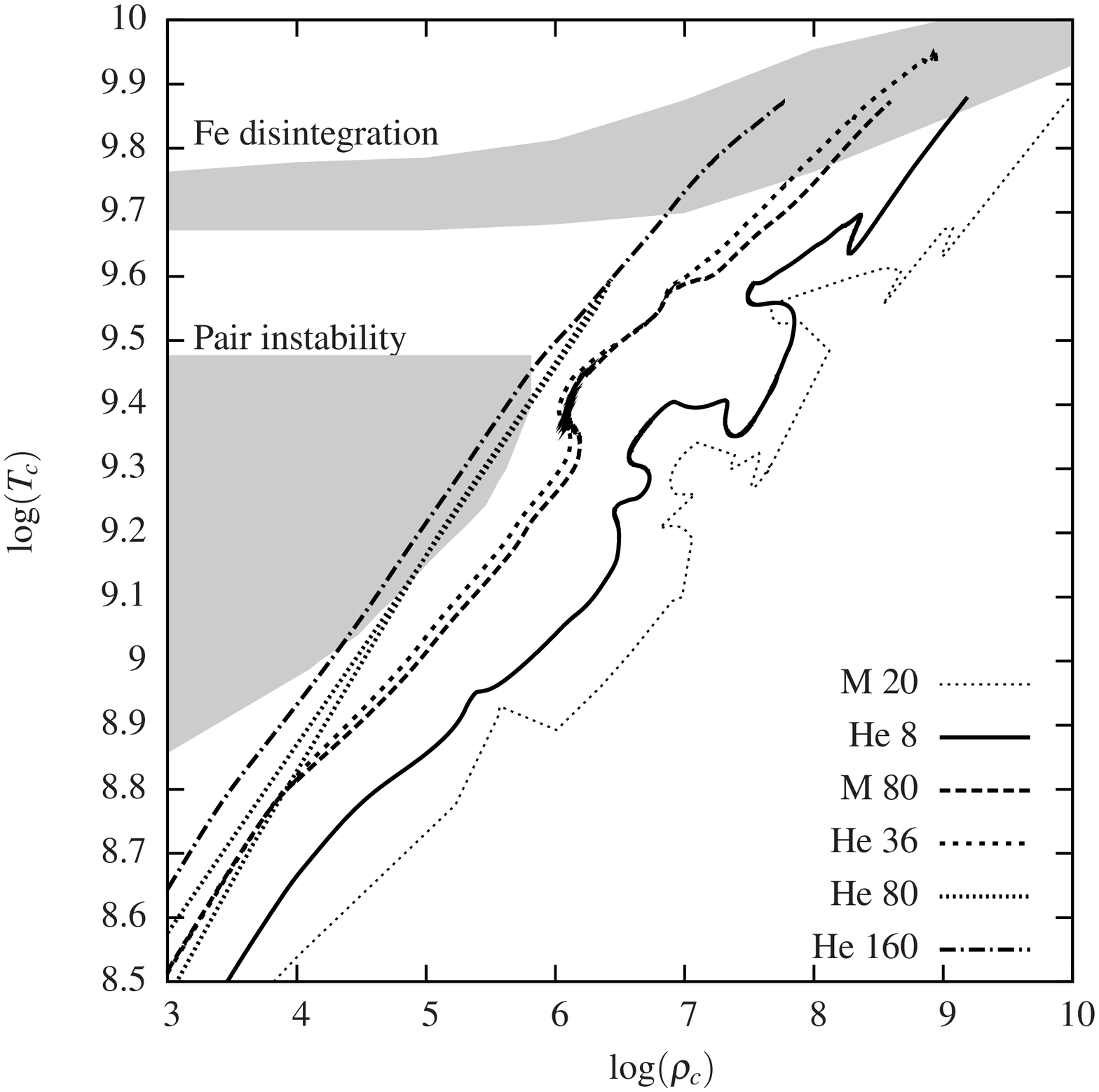}
\end{minipage}
\begin{minipage}[t]{0.5\linewidth}
\centering
\includegraphics[width=1.\linewidth]{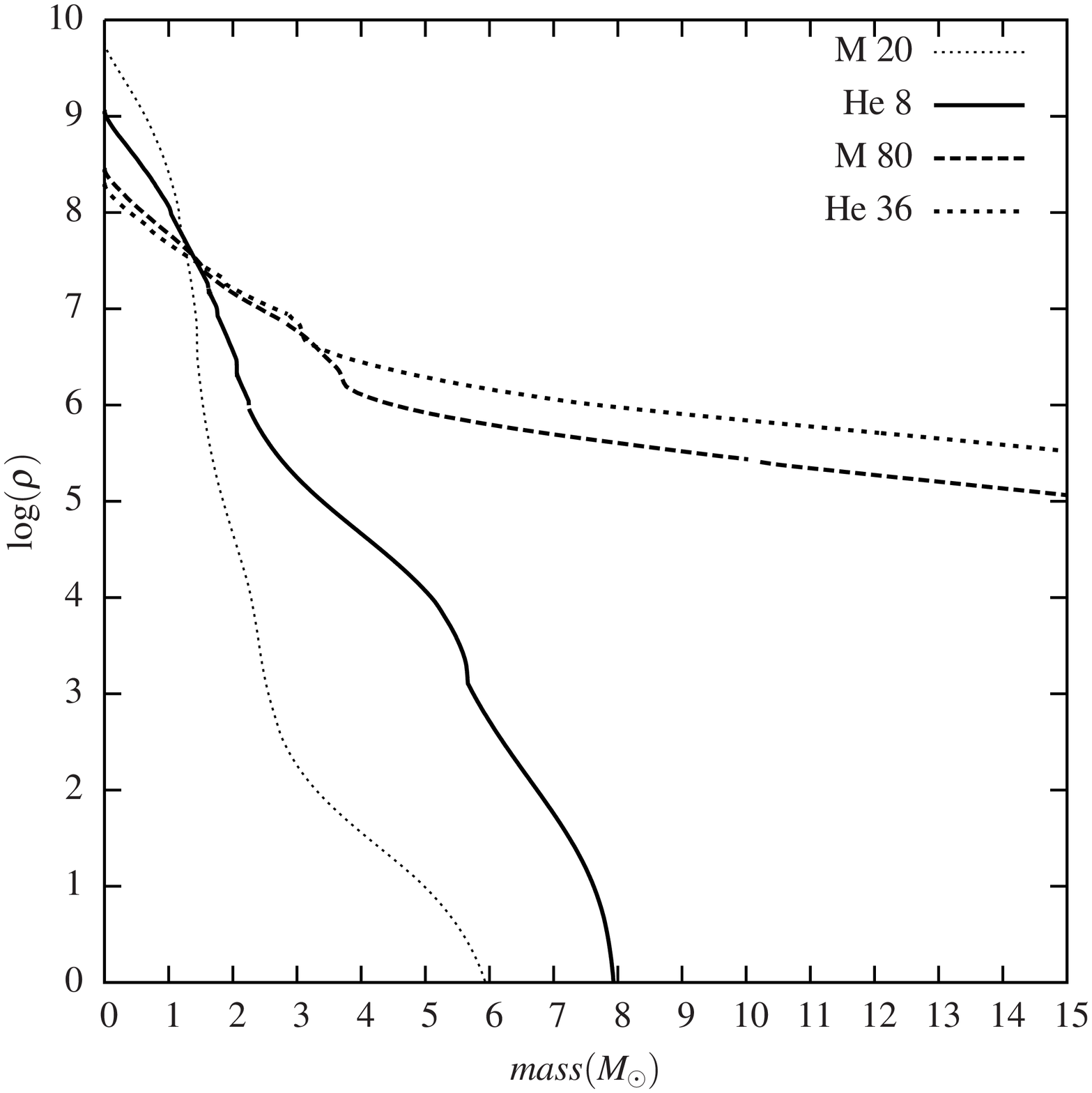}
\end{minipage}
\caption{Evolution of the central density and temperature (left
panel) and pre-SN density structure (right panel). Each line is
labeled ``M'' for stellar models and ``He'' for He-core models,
followed by the mass of the model.} \label{fig:rhoc_tc}
\end{figure}

\begin{figure}
\begin{minipage}[t]{0.49\linewidth}
\centering
\includegraphics[width=1.\linewidth]{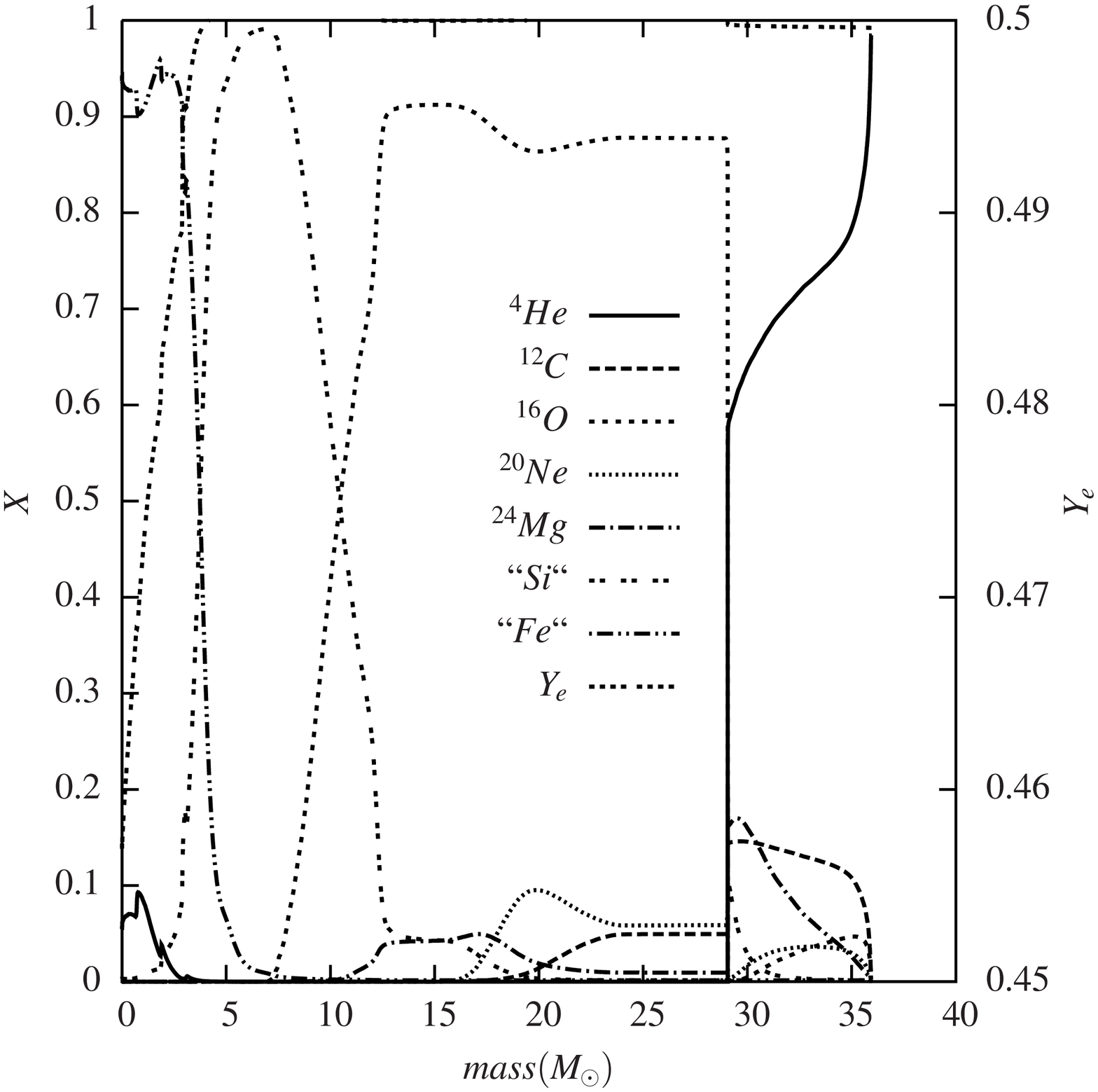}
\caption{Pre-SN composition of the He36 model. ``Si'' and ``Fe''
stand for the total of Si- and Fe-group elements respectively.}
\label{fig:He36}
\end{minipage}
\hspace{0.02\linewidth}
\begin{minipage}[t]{0.49\linewidth}
\centering
\includegraphics[width=1.\linewidth]{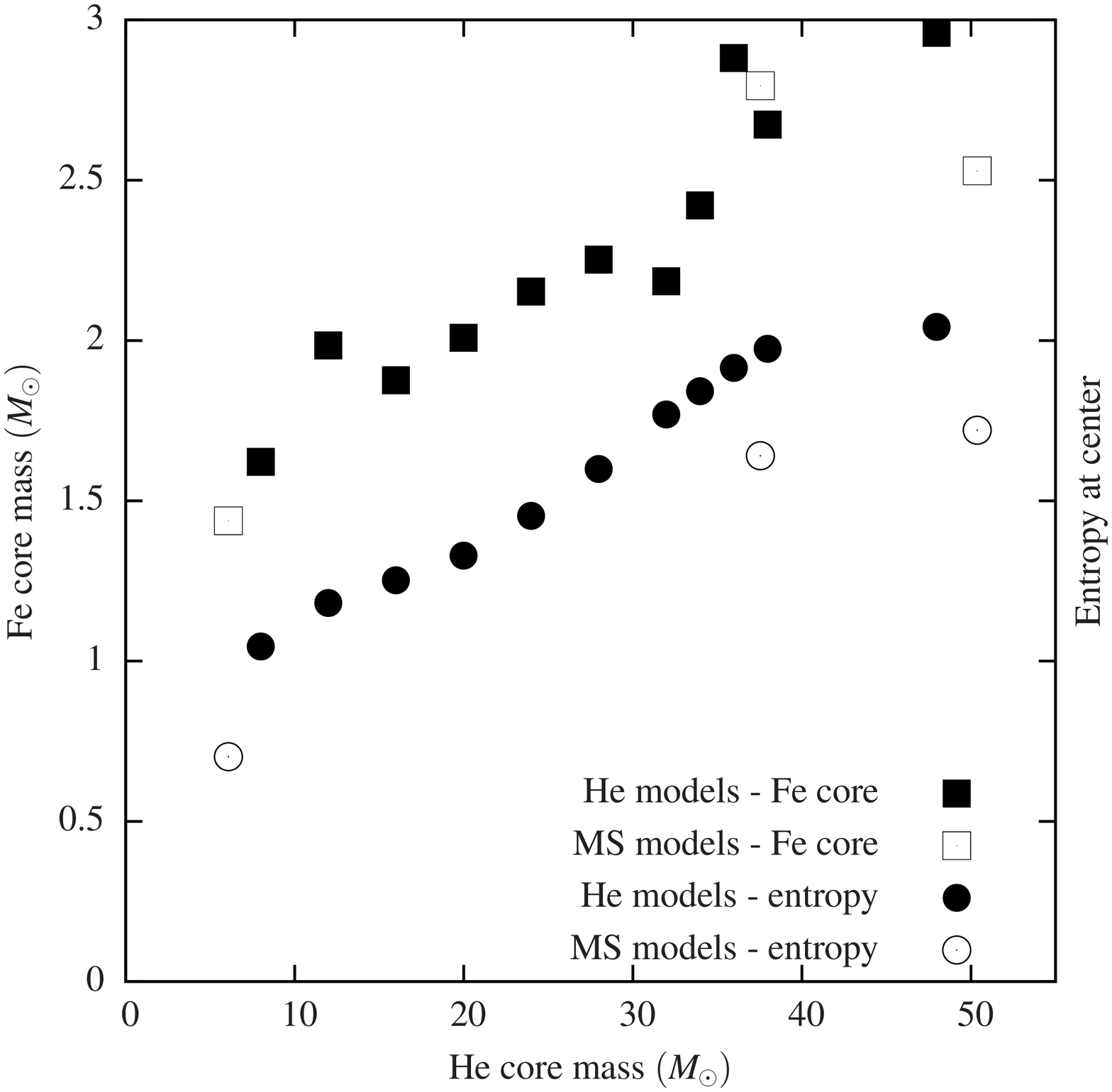}
\caption{Mass of the Fe-core (squares) and central entropy per
baryon (circles) for the computed models. Filled shapes designate
He-core models, open shapes - stellar models.} \label{fig:fe_cores}
\end{minipage}

\end{figure}

Among the He-core models we computed, those in the mass range $M
\leq 36\,M_\odot$ do not reach pair instability and end their lives
as CCSN. Fig. \ref{fig:rhoc_tc} shows the evolution of the central
density and temperature (left panel) and the density structure of
the pre-SN (right panel), at the moment when the central temperature
reaches $7 \times 10^{9}\,K$. The two extreme models He8 and He36
are shown, as well as M80 which has a He-core mass similar to the
He36 model, and M20 - a typical CCSN progenitor. The left panel also
shows two He-core models - He80 and He160 that reach pair
instability. Note that the two models He36 and M80 are indeed very
close to each other. An example of the composition at the pre-SN
stage ($T_c=7 \times 10^9 K$) is shown in Fig. \ref{fig:He36}. The
Fe-core mass is $\approx 3\,M_\odot$, topped by a shell of $\approx
10\,M_\odot$ of Si-group elements. The size of the Fe-core (defined
as the mass coordinate where the electron mole fraction $Y_e<0.49$)
and the central entropy per baryon at the pre-SN stage for the whole
set of models is shown in Fig. \ref{fig:fe_cores}. Note that the
size of the Fe-core is slightly non-monotonic. The central entropy,
is monotonic with mass, but slightly differs between He-core and
stellar models.

As can be seen in the above figures, the outstanding features of
these massive models compared with the typical CCSN example M20 are:
\begin{enumerate}
  \item Lower central density and higher central entropy.
  \item A much shallower density profile.
  \item Relatively large Fe-cores, up to about $3\,M_\odot$, and a
  large amount (up to about $10\,M_\odot$) of Si-group elements.
\end{enumerate}

These differences might have a considerable impact on the behavior
of these models during core collapse and on the outcome of the
explosion, a question which we hope to address in the future.

\begin{acknowledgments}
I would like to thank Zalman Barkat for many hours of fruitful
discussion. I would like to acknowledge David Arnett for providing
his TYCHO code for public use.
\end{acknowledgments}

\bibliography{Mybib}

\end{document}